\documentclass[12pt, onecolumn]{IEEEtran}
\usepackage{amsmath}
\usepackage{amssymb}
\usepackage{mathrsfs}
\usepackage{cite}
\usepackage{epsfig}
\usepackage{epsf}
\usepackage{theorem}
\usepackage{graphics}
\usepackage[active]{srcltx}

\renewcommand{\baselinestretch}{1.9}

\begin{document}
\title{ An Efficient Adaptive Distributed Space-Time Coding Scheme for Cooperative Relaying \\
\thanks{$^{*}$ This work is financially supported by Nortel Networks and the corresponding matching funds by the Natural Sciences
and Engineering Research Council of Canada (NSERC), and Ontario Centers of Excellence (OCE).}
\thanks{$^{*}$
The material in this paper was presented in part at the 42nd Conference on IEEE
Information Sciences and Systems (CISS), Princeton University,
Princeton, NJ, March 19-21, 2008 \cite{JamshidCISS08}.}}

\author{\small \textbf{Jamshid Abouei, Hossien Bagheri, and Amir K. Khandani} \\
\small Coding and Signal Transmission Laboratory (www.cst.uwaterloo.ca)\\
Department of Electrical and Computer Engineering, University of Waterloo\\
Waterloo, Ontario, Canada, N2L 3G1 \\
Tel: 519-884-8552, Fax: 519-888-4338\\
Emails: \{jabouei, hbagheri, khandani\}@cst.uwaterloo.ca}

\maketitle


\begin{abstract}
A non-regenerative dual-hop wireless system based on a distributed
space-time coding strategy is considered. It is assumed that each
relay retransmits an appropriately scaled space-time coded version
of its received signal. The main goal of this paper is to investigate
a power allocation strategy in relay stations, which is based on minimizing
the outage probability. In the high signal-to-noise ratio regime
for the relay-destination link, it is shown that a threshold-based
power allocation scheme (i.e., the relay remains silent if its channel
gain with the source is less than a prespecified threshold) is optimum.
Monte-Carlo simulations show that the derived on-off power allocation
scheme performs close to optimum for finite signal-to-noise ratio values.
Numerical results demonstrate a dramatic improvement in system
performance as compared to the case that the relay
stations forward their received signals with full power. In addition,
a hybrid amplify-and-forward/detect-and-forward scheme is proposed for
the case that the quality of the source-relay link is good. Finally,
the robustness of the proposed scheme in the presence of channel
estimation errors is numerically evaluated.
\end{abstract}


\section{Introduction}

A primary challenge in wireless networks is to mitigate the effect
of the multipath fading. Exploiting techniques such as time, frequency
and space diversity are the most effective methods to combat the
channel fading. Due to the dramatic reduction in the size of wireless
devices along with the power and the cost limitations, it is not
practical to install multiple antennas on mobile stations. A heuristic
solution to this problem is to use a collection of distributed antennas
belonging to multiple users
\cite{Sendonaris1ITC1103, Sendonaris2ITC1103, LanemanITIT1204}. This
new diversity scheme, referred to as cooperative diversity (also
known as \textit{cooperative relaying}), has attracted considerable
attention in ad hoc and sensor wireless networks in recent years. In
particular, this approach is considered as an option in many wireless
systems; including IEEE 802.16j mobile multihop relay-based (MMR)
networks. Unlike conventional relaying systems that forward the received
signal in a relay chain, cooperative relaying takes one step
further; i.e., multiple relay nodes work together to
achieve a better performance.

Cooperative relay networks have been addressed from different
perspectives; including capacity and outage probability analysis
\cite{Sendonaris1ITC1103, Sendonaris2ITC1103, LanemanITIT1003,
LanemanITIT1204, HasnaITWC0604, MadsenITIT0605}, resource allocation
(e.g., power allocation \cite{ZhaoITWC0807}), coding
\cite{KramerITIT0905}, relay selection \cite{LanemanITIT1204,
BletsasJSAC0306}, etc. Central to the study of cooperative relay
systems is the problem of using \textit{distributed space-time
coding} (DSTC) technique as well as efficient power allocation
schemes in regenerative and non-regenerative configurations
\cite{LanemanITIT1003, NabarJSAC0804, ChangICASSP04, MiyanoVTC2004,
ScutariITWC0905, AnghelKavehITWC0306, RongZhangVTC2006, HeICC2007, GuoITWC0508}.

The first study on using DSTC scheme in cooperative relay networks
was framed in \cite{LanemanITIT1003} where several relay nodes transmit
jointly to the same receiver in order to achieve full spatial
diversity in terms of the number of cooperating nodes. Reference
\cite{LanemanITIT1003} analyzes the outage capacity in the high signal-to-noise ratio (SNR)
regime. Nabar \textit{et al.} \cite{NabarJSAC0804} analyze the
pairwise error probability of an amplify-and-forward (AF) single-relay
system which relies on a DSTC scheme.
In \cite{ChangICASSP04}, the authors investigate the high
SNR uncoded bit-error rate (BER) of a two-relay system using a switching
scheme for QPSK modulation. Scutari and Barbarossa \cite{ScutariITWC0905} analyze the
performance of regenerative relay networks with DSTC where the error
on the source-relay link propagates into the second phase of the
transmission in the relay-destination link. They show how to allocate
the transmission power between source and relay to minimize the average
BER. The BER of single and dual-hop non-regenerative relay
systems with DSTC has been analyzed in \cite{AnghelKavehITWC0306}
where different transmission policies are investigated in order to
maximize the end-to-end SNR. The BER of regenerative relay networks
using the DSTC scheme, along with different power allocation
strategies over non-identical Ricean channels is analyzed in
\cite{HeICC2007}. Zhao \textit{et al.} \cite{ZhaoITWC0807} introduce
two power allocation algorithms for AF relay networks
to minimize the outage probability without the DSTC scheme.

In this paper, we consider a dual-hop wireless system consisting
of a source, two parallel relay stations (RS) and a destination.
Assuming that each relay knows its channel with the source, the RSs
collaborate with each other by transmitting a space-time coded (STC)
version of their received signals to the destination. In such configuration,
if the instantaneous received SNR of the relays are unbalanced,
the performance of the system degrades substantially. To overcome
this problem, an adaptive distributed space-time coding (ADSTC)
method is proposed, in which instead of transmitting the noisy
signal at each relay with full power, the RS retransmits an
appropriately scaled STC version of the received signal. The main
goal is to optimize the scaling factor based on the channel-state
information (CSI) available at each relay to minimize the outage
probability.

The above scheme is different from the power allocation algorithms
studied in \cite{ZhaoITWC0807}; primarily we utilize an ADSTC scheme
based on the CSI available at each relay, while in \cite{ZhaoITWC0807},
no DSTC scheme is used. In our
scheme, no information is exchanged between the relay stations. This
is to be contrasted with the algorithms proposed in
\cite{AnghelKavehITWC0306}, in which the relays need to exchange
some information with each other in order to establish which relay transmits.
In addition, our scheme is different from the power allocation
strategies in \cite{ZhangEURASIP2008, ChenGLOBECOM2005, LoVTC2007},
in which the best relays are selected based on the feedback from the destination.

In the high SNR regime for the relay-destination link, it is shown that
a threshold-based power allocation scheme (i.e., the relay remains silent
if its channel gain with the source is less than a prespecified threshold)
is optimum. Numerical results indicate that the derived on-off power allocation
scheme provides a significant performance improvement as compared to the case that the RSs
forward their received signals with full power all the time.
To further improve the system performance when the quality of the source-relay link is good,
a hybrid amplify-and-forward/detect-and-forward scheme is proposed.

In the proposed scheme, we assume that perfect channel knowledge
is available at the receiver. In practice, however, the performance of
the system is degraded due to the channel estimation error. In the
last part of this paper, we address this issue and evaluate the
robustness of the proposed scheme in the presence of channel estimation
errors.

The rest of the paper is organized as follows. In Section
\ref{model}, the system model and objectives are described. The
performance of the system is analyzed in Section \ref{analysis}. In
Section \ref{simulation}, the simulation results are presented. The
effect of channel estimation errors on the proposed scheme is
evaluated in Section \ref{robust}. Finally, conclusions are drawn in
Section \ref{conclusion1}.

\textit{Notations:} Throughout the paper, we use boldface capital and lower case
letters to denote vectors and matrices, respectively. $\Vert
~\textbf{a}~ \Vert$ indicates the Euclidean norm of the vector
$\textbf{a}$. The conjugate and conjugated transposition of a
complex matrix $\textbf{A}$ are denoted by $\textbf{A}^{*}$ and
$\textbf{A}^{\dagger}$, respectively. The $n\times n$ identity
matrix is denoted by $\textbf{I}_{n}$. A circularly symmetric
complex Gaussian random variable (r.v.) is represented by $Z=X+jY
\sim \mathcal{CN}(0,\sigma^{2})$, where $X$ and $Y$ are independent
and identically distributed (i.i.d.) normal r.v.'s with
$\mathcal{N}(0,\frac{\sigma^{2}}{2})$. Also, $A \rightarrow B$
represents the link from node $A$ to node $B$. Finally, $\mathbb{P} \{. \}$
and $\mathbb{E}[.]$ denote the probability of the given event and the
expectation operator, respectively.

\section{System Model and Objectives}\label{model}

In this work, we consider a cooperative relay system consisting of
one base station (BS), multiple relay stations randomly located within
the network region, and multiple mobile stations (MSs) (Fig. \ref{fig: model}-a).
All the nodes are assumed to have a single antenna. The relay stations are assumed to be
non-regenerative, i.e., they perform some simple operations on the
received signals and forward them to the MSs \cite{AnghelKavehITWC0306}. Also, it is
assumed that no information is exchanged between the relays. In this
set-up, each pair of the relay stations cooperate with each other to
forward their received signals from the source to the
destination\footnote{In the sequel and for the sake of simplicity of
notations, the BS and the MS are denoted by the source (S) and the destination (D), respectively.}
(Fig. \ref{fig: model}-b).

The channel model is assumed to be frequency-flat block Rayleigh
fading with the path loss effect. Let us denote the channel coefficients of the
links $\textrm{S} \rightarrow \textrm{RS}_{i}$ and RS$_{i}$
$\rightarrow$ D as $\mathcal{L}_{sr_{i}}$ and
$\mathcal{L}_{r_{i}d}$, $i=1,2$, respectively. In this case, the channel gain
between S and RS$_{i}$ is represented by $g_{sr_{i}} \triangleq
\vert \mathcal{L}_{sr_{i}}\vert^{2} =\Gamma_{sr_{i}}
\vert h_{sr_{i}} \vert^{2}$, where $\Gamma_{sr_{i}} \triangleq 10^{-\Upsilon_{sr_{i}}/10}$ is the
gain associated with the path loss $\Upsilon_{sr_{i}}$ [dB], and the complex
random variable $h_{sr_{i}}$ is the fading channel coefficient.
Similarly, the channel gain of the link $\textrm{RS}_{i} \rightarrow
\textrm{D}$ is represented by $g_{r_{i}d} \triangleq \vert
\mathcal{L}_{r_{i}d}\vert^{2}=\Gamma_{r_{i}d}
\vert h_{r_{i}d} \vert^{2}$, where $\Gamma_{r_{i}d} \triangleq 10^{-\Upsilon_{r_{i}d}/10}$ and
$h_{r_{i}d}$ are the gain associated with the path loss $\Upsilon_{r_{i}d}$ [dB] and the channel fading
coefficient, respectively. Under the Rayleigh fading channel model,
$\vert h_{sr_{i}} \vert^{2}$ and $\vert h_{r_{i}d} \vert^{2}$
are exponentially distributed with unit mean (and unit variance). Also, the
background noise at each receiver is assumed to be additive white
Gaussian noise (AWGN).
For this model, communication between S and D is
performed based on the following steps:
\begin{figure}
\centerline{\psfig{figure=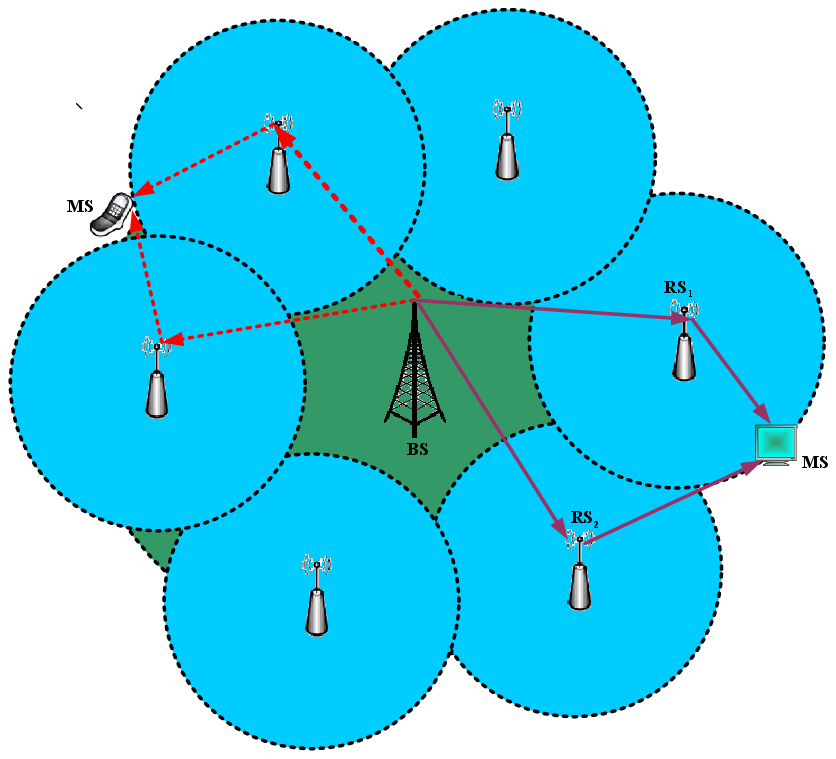,width=4.6in}}
\vspace{-7pt} \center{\hspace{16pt} \small{(a)}} \vspace{10pt}
\hspace{1pt} \centerline{\psfig{figure=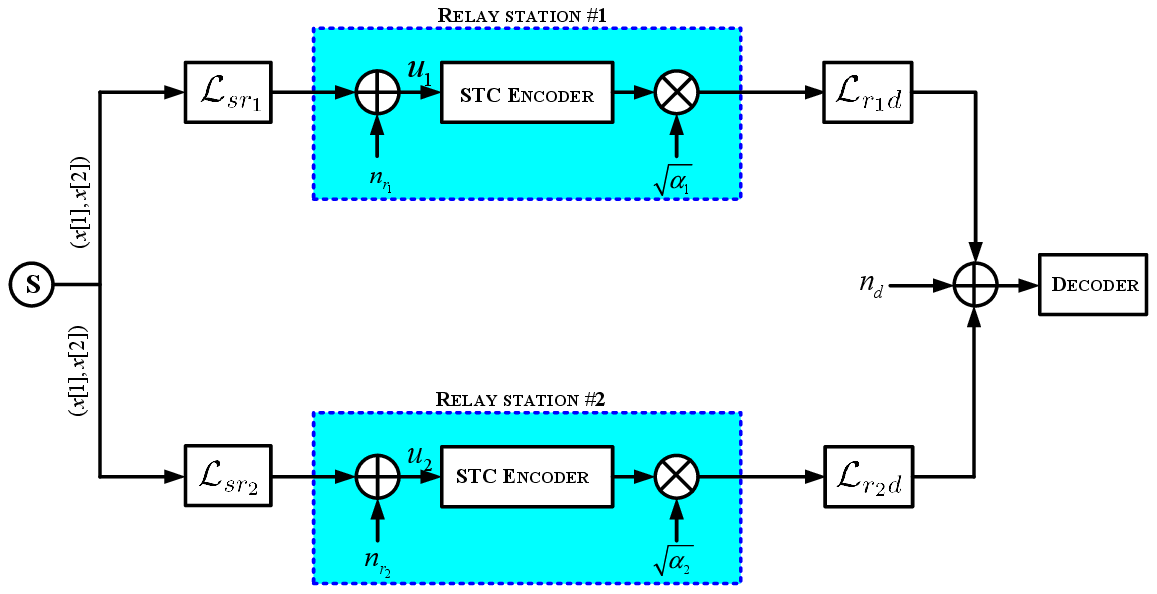,width=4.3in}}
\vspace{-16pt}
\center{\hspace{14pt} \small{(b)}} \\
\vspace{-7pt} \caption[a) $\lambda_{\ell}$ and b) for.] {\small{a)
Cooperative relay network, and  b) a discrete-time baseband equivalent model of a dual-hop wireless system.}}
\label{fig: model}
\end{figure}
\underline{ \textbf{i) Data Transmission:}} This is performed in two phases
and through two hops. In the first phase, the source broadcasts the symbols
$x[1]$ and $x[2]$ (normalized to have unit energy) to the RSs
in two consecutive time slots and over one frequency
band. The received discrete-time baseband signal at RS$_{i}$ is given by
\begin{equation}\label{eqn: 01}
u_{i}[k]= \sqrt{\textrm{P}_{s}}\mathcal{L}_{sr_{i}}  x[k]+n_{r_{i}}[k],~~i=1,2,
\end{equation}
where $k=1,2$, represents the transmission time index,
$\textrm{P}_{s}$ is the transmit power of the source
and $n_{r_{i}}[k] \sim \mathcal{CN}(0,N_{0})$ is the background
noise at RS$_{i}$. The channel gains are assumed to remain constant over two
successive symbol transmissions. Also, we assume that RS$_{i}$
has perfect information about $h_{sr_{i}}$. To satisfy the
power constraints at the RSs, we normalize
$u_{i}[k]$ to $\sqrt{\mathbb{E}_{x}\left[\vert u_{i}[k] \vert ^{2}
\Big\vert h_{sr_{i}} \right]},~i=1,2~$ \cite{ZhaoITWC0807, RongZhangVTC2006, RibeiroITWC0505}. Thus, (\ref{eqn: 01})
can be written as
\begin{equation}\label{eqn: 02}
y_{i}[k] = \dfrac{1}{\sqrt{\textrm{P}_{s}g_{sr_{i}}+N_{0}}}\left (\sqrt{\textrm{P}_{s}}\mathcal{L}_{sr_{i}} x[k]+n_{r_{i}}[k]\right),
\end{equation}
for $i,k=1,2$. Using the fact that $x[k]$ is independent of $n_{r_{i}}[k]$, it yields
$\mathbb{E}\left[\vert y_{i}[k] \vert ^{2}\Big\vert h_{sr_{i}} \right]=1$.

\begin{table}
   \label{table1}
\caption{Transmission Policy at RSs}
\centering
  \begin{tabular}{|c|c|c|}
  \hline
   & RS$_{1}$  & RS$_{2}$  \\
  \hline
   $1^{st}$ time slot &  $\sqrt{\alpha_{1}} e^{-j\theta_{sr_{1}}}y_{1}[1]$ &  $\sqrt{\alpha_{2}}e^{-j\theta_{sr_{2}}} y_{2}[2]$\\
  \hline
   $2^{nd}$ time slot &  $-\sqrt{\alpha_{1}} \left (e^{-j\theta_{sr_{1}}}y_{1}[2] \right)^{*}$ &  $\sqrt{\alpha_{2}}\left(e^{-j\theta_{sr_{2}}} y_{2}[1]\right)^{*}$\\
  \hline
  \end{tabular}
\end{table}
In the second phase, the RSs cooperate with each other and
forward the space-time coded version of their received noisy signals
to the destination over the same or another frequency band\footnote{It is
assumed that RSs use the same transmission protocol based on the IEEE 802.16j time division
duplex (TDD) frame structure \cite{IEEE802.16j-doc}.} (Table I). In such configuration, RS$_{i}$
multiplies $y_{i}[.]$ by the scaling factor $\sqrt{\alpha_{i}}$,
where $0 \leq \alpha_{i} \leq 1$. It should be noted that the
phase of S $\rightarrow$ RS$_{i}$ link, denoted by
$\theta_{sr_{i}}$, is compensated at RS$_{i}$ through multiplying
the received signal by the factor
$e^{-j\theta_{sr_{i}}}=\frac{h_{sr_{i}}^{*}}{\vert h_{sr_{i}} \vert}$
\cite{AnghelKavehITWC0306, RongZhangVTC2006} \footnote{We will explain the main reason
for using the phase compensation shortly.}.
In general, due to the different distances between each RS and D,
the arrival time of the received signals at D may be different.
The cyclic prefix added to the orthogonal frequency-division multiplexing
(OFDM) symbols mitigates the effect of
the time delay, and it preserves the orthogonality of the tones.
Thus, we can apply the above scheme in each tone.

In each time slot, D receives a superposition of the transmitted signals by RSs.
To this end, the received signals at D in the first and the second time slots are given by
\begin{eqnarray}
\label{eqn: 003} r_{d}[1] & = & \sqrt{\alpha_{1}} e^{-j\theta_{sr_{1}}}y_{1}[1]\sqrt{\textrm{P}_{r_{1}}}\mathcal{L}_{r_{1}d}+
 \sqrt{\alpha_{2}}e^{-j\theta_{sr_{2}}}y_{2}[2]\sqrt{\textrm{P}_{r_{2}}}\mathcal{L}_{r_{2}d}+n_{d}[1] ,\\
\label{eqn: 004} r_{d}[2] & = & -\sqrt{\alpha_{1}} \left (e^{-j\theta_{sr_{1}}}y_{1}[2] \right)^{*}\sqrt{\textrm{P}_{r_{1}}}\mathcal{L}_{r_{1}d}+
\sqrt{\alpha_{2}}\left(e^{-j\theta_{sr_{2}}} y_{2}[1]\right)^{*}\sqrt{\textrm{P}_{r_{2}}}\mathcal{L}_{r_{2}d}+n_{d}[2],
\end{eqnarray}
respectively, where $\textrm{P}_{r_{i}}$ is the transmit power of RS$_{i}$ and
$n_{d}[k] \sim \mathcal{CN}(0,N_{0})$ represents
the background noise at the destination. It should be noted that due
to the large distance between S and D (or due to the strong shadowing),
we ignore the received signal of the direct link S $\rightarrow$ D.
Substituting (\ref{eqn: 02}) in (\ref{eqn: 003}) and (\ref{eqn: 004}) yields
\begin{eqnarray}
\label{eqn05} r_{d}[1] & = & \mathcal{L}^{'}_{1}x[1]+\mathcal{L}^{'}_{2}x[2]+z_{d}[1] ,\\
\label{eqn06} r_{d}[2] & = & \mathcal{L}_{2}^{'}x^{*}[1]- \mathcal{L}_{1}^{'}x^{*}[2]+z_{d}[2],
\end{eqnarray}
where
\begin{equation}\label{eqn: 03}
\mathcal{L}^{'}_{i} \triangleq \mathcal{L}_{r_{i}d}\sqrt{\dfrac{\alpha_{i}\textrm{P}_{s}\textrm{P}_{r_{i}}g_{sr_{i}}}{\textrm{P}_{s}g_{sr_{i}}+N_{0}}},~~i=1,2,
\end{equation}
and
\begin{eqnarray}
z_{d}[1]& \triangleq &\dfrac{\mathcal{L}^{'}_{1}}{\sqrt{\textrm{P}_{s}} \mathcal{L}_{sr_{1}}} n_{r_{1}}[1]+\dfrac{\mathcal{L}^{'}_{2}}{\sqrt{\textrm{P}_{s}} \mathcal{L}_{sr_{2}}}n_{r_{2}}[2]+n_{d}[1], \\
z_{d}[2]& \triangleq &-\dfrac{\mathcal{L}^{'}_{1}}{\sqrt{\textrm{P}_{s}} \mathcal{L}^{*}_{sr_{1}}}n^{*}_{r_{1}}[2]+\dfrac{\mathcal{L}^{'}_{2} }{\sqrt{\textrm{P}_{s}} \mathcal{L}^{*}_{sr_{2}}}n^{*}_{r_{2}}[1]+n_{d}[2].
\end{eqnarray}
It is observed that $z_{d}[1]$ is independent of $z_{d}[2]$ and $\mathbb{E} \Big[\vert z_{d}[1] \vert^{2} \Big \vert \textbf{h} \Big]=\mathbb{E} \Big[\vert z_{d}[2] \vert^{2} \Big \vert \textbf{h} \Big] = \sigma^{2}$, where $\textbf{h}\triangleq[h_{sr_{1}},h_{sr_{2}},h_{r_{1}d},h_{r_{2}d}]$ and
\begin{equation}\label{eqn: 04}
\sigma^{2} \triangleq \left (\dfrac{\vert \mathcal{L}^{'}_{1} \vert ^{2}}{g_{sr_{1}}}+\dfrac{\vert \mathcal{L}^{'}_{2} \vert ^{2}}{g_{sr_{2}}}\right)\dfrac{N_{0}}{\textrm{P}_{s}}+N_{0}.
\end{equation}

\underline{ \textbf{ii) Decoding Process:}} Amplify-and-forward relay networks require full
CSI at the destination to coherently decode the received signals. The required
channel information can be acquired by transmitting pilot signals\footnote{ Channel estimation can be done in two steps. In the first step, RS$_{i}$ transmits a pilot
sequence to provide $\mathcal{L}_{r_{i}d}$ to D. In the second step, S transmits another pilot sequence
to provide $\mathcal{L}_{sr_{i}}$  to RS$_{i}$ and $\mathcal{L}_{sr_{i}}\mathcal{L}_{r_{i}d}$
to D. In this step, RSs work as usual.}.
According to the Alamouti scheme \cite{AlamoutiJSAC0898}, we have
\begin{displaymath}
\left [\begin{array}{c}r_{d}[1]\\
r^{*}_{d}[2]\end{array} \right]
=  \left [ \begin{array}{cc} \mathcal{L}^{'}_{1} & \mathcal{L}^{'}_{2}\\
\mathcal{L}_{2}^{'*}& - \mathcal{L}_{1}^{'*}\end{array} \right]
 \left [\begin{array}{c}x[1]\\
x[2]\end{array} \right]
+\left[ \begin{array}{c}z_{d}[1] \\
z^{*}_{d}[2] \end{array}\right],
\end{displaymath}
or equivalently $\textbf{r}_{d}=\boldsymbol{\mathcal{L}}\textbf{x}+\textbf{z}_{d}$. It is worth
mentioning that due to the conjugate terms in (\ref{eqn: 004}), the heuristic phase compensation
at RSs (i.e., $e^{-j\theta_{sr_{i}}}=\frac{h_{sr_{i}}^{*}}{\vert h_{sr_{i}} \vert}$)
is done to preserve the orthogonal property of the processing matrix $\boldsymbol{\mathcal{L}}$, i.e.,
$\boldsymbol{\mathcal{L}}^{\dagger}\boldsymbol{\mathcal{L}}=\left( \vert \mathcal{L}_{1}^{'} \vert ^{2}+ \vert \mathcal{L}^{'}_{2} \vert ^{2} \right)\textbf{I}_{2} \triangleq \Lambda \textbf{I}_{2}$. Thus, we can use the Alamouti scheme. Also, upon this phase
compension assumption and the AF scheme at the relay nodes, the above scenario is converted to the original STC
problem. Hence, the mentioned decoder is maximum likelihood (ML) optimal. For this configuration,
the input of the ML decoder is given by
\cite{TarokhITIT0799}
\begin{eqnarray}
\mathbf{\tilde{r}}_{d} & \triangleq & \boldsymbol{\mathcal{L}}^{\dagger}\mathbf{r}_{d}\\
\label{eqn: ML01}& = & \Lambda \mathbf{x}+\mathbf{\tilde{z}}_{d},
\end{eqnarray}
where $\mathbf{\tilde{z}}_{d}
\triangleq \boldsymbol{\mathcal{L}}^{\dagger}\mathbf{z}_{d}$. In
this case, the two-dimensional decision rule used in the ML decoder
is
\begin{equation}
\mathbf{\hat{x}}=\arg~\min_{\mathbf{\hat{x}}\in \mathbb{S}^{2}} \Vert \mathbf{\tilde{r}}_{d}- \Lambda  \mathbf{\hat{x}} \Vert ^{2},
\end{equation}
where $\mathbb{S}^{2}$ denotes the corresponding signal constellation set.

Let us denote the average signal-to-noise ratios of S $ \rightarrow$ RS$_{i}$ and
RS$_{i}$ $\rightarrow$ D by SNR$_{sr_{i}} \triangleq \frac{\textrm{P}_{s}}{N_{0}}\Gamma_{sr_{i}}$
and SNR$_{r_{i}d}\triangleq \frac{\textrm{P}_{r_{i}}}{N_{0}}\Gamma_{r_{i}d}$, $i=1,2$,
respectively. Here, we assume that SNR$_{sr_{i}}$ and SNR$_{r_{i}d}$ are known at RS$_{i}$.
Assuming $h_{sr_{i}}$ is available at RS$_{i}$, the main objective is to select
the optimum factor $\alpha_{i}$ such that the outage probability is minimized.

\section{Performance Analysis} \label{analysis}
In this section, we characterize the performance of the model described in Section
\ref{model} in terms of the outage probability. For this purpose, we first obtain the
instantaneous end-to-end SNR. Then, we derive an expression for the outage probability
in the high SNR$_{r_{i}d}$ regime.

Since, matrix $\boldsymbol{\mathcal{L}} \vert \textbf{h}$ is deterministic and noting that
$\mathbb{E}\left[\mathbf{z}_{d}\mathbf{z}_{d}^{\dagger} \vert
\textbf{h}\right]=\sigma^{2}\textbf{I}_{2}$, we conclude from (\ref{eqn: ML01}) that
\begin{equation}
\mathbb{E}\left[\mathbf{\tilde{z}}_{d}\mathbf{\tilde{z}}_{d}^{\dagger} \vert \textbf{h} \right]=\mathbb{E}\left[\boldsymbol{\mathcal{L}}^{\dagger}\textbf{z}_{d}\mathbf{z}_{d}^{\dagger}\boldsymbol{\mathcal{L}} \vert \textbf{h} \right]=\Lambda \sigma^{2}\textbf{I}_{2}.
\end{equation}
In fact, the noise vector $\mathbf{\tilde{z}}_{d} \vert \textbf{h}$
is a white Gaussian noise. Denoting $\gamma$ as the instantaneous
end-to-end SNR and using (\ref{eqn: ML01}), we have
\begin{equation}\label{eqn: 05}
\gamma  =  \dfrac{\Lambda^{2}}{\Lambda \sigma^{2}}= \dfrac{\vert \mathcal{L}^{'}_{1} \vert ^{2}+\vert \mathcal{L}^{'}_{2} \vert ^{2}}{\left (\dfrac{\vert \mathcal{L}^{'}_{1} \vert ^{2}}{g_{sr_{1}}}+\dfrac{\vert \mathcal{L}^{'}_{2} \vert ^{2}}{g_{sr_{2}}}\right)\dfrac{N_{0}}{\textrm{P}_{s}}+N_{0}}.
\end{equation}

In the outage-based transmission framework, the outage occurs whenever
$\gamma$ is less than a threshold $\gamma_{t} >0$. In this case, the
outage probability can be expressed as
\begin{equation}\label{eqn: 06}
\textrm{P}_{out} \triangleq \mathbb{P} \lbrace \gamma < \gamma_{t} \rbrace.
\end{equation}

The goal is to minimize the objective function defined in
(\ref{eqn: 06}) subject to $0 \leq \alpha_{i} \leq 1$, $i=1,2$.
The global optimal solution can be obtained through
centralized algorithms. This is in contrast to the distributed
scheme proposed in Section \ref{model}. Since, it is difficult to directly
compute the exact expression for the outage probability, we first
present the outage probability in the high SNR$_{r_{i}d}$ regime. Then, we propose an optimum
distributed power allocation strategy for the RS$_{i}$ to minimize
the objective function in (\ref{eqn: 06}).

Defining $\textbf{g} \triangleq \Big[\vert h_{sr_{1}}\vert ^{2}=v_{1}, \vert h_{sr_{2}}\vert ^{2}=v_{2} \Big]$, and
using the fact that the link S $ \rightarrow$ RS$_{1}$ is independent of the link S $ \rightarrow$ RS$_{2}$,
the outage probability is given by
\begin{equation}\label{eqn: 07}
\textrm{P}_{out} = \mathbb{E}_{\textbf{g}}[\Omega(\textbf{g})]=\int_{0}^{\infty} \int_{0}^{\infty} \Omega(\textbf{g}) e^{-v_{1}}e^{-v_{2}}dv_{1}dv_{2},
\end{equation}
where $\Omega(\textbf{g}) \triangleq \mathbb{P} \lbrace \gamma < \gamma_{t} \vert \textbf{g} \rbrace$.
It is concluded from (\ref{eqn: 03}) and (\ref{eqn: 05}) that
\begin{equation}\label{eqn: 08}
\Omega(\textbf{g}) = \mathbb{P} \left. \left \lbrace  \dfrac{X_{1} v_{1}+X_{2} v_{2}}{(\dfrac{X_{1}}{\Gamma_{sr_{1}}}+\dfrac{X_{2}}{\Gamma_{sr_{2}}})\dfrac{N_{0}}{\textrm{P}_{s}}+N_{0}} <\gamma_{t} \right \vert \textbf{g} \right \rbrace,
\end{equation}
where $X_{i}\triangleq \mathscr{F}(v_{i})\vert h_{r_{i}d}\vert ^{2}$ with
\begin{equation}\label{eq: 09}
 \mathscr{F}(v_{i}) \triangleq \dfrac{\alpha_{i}\textrm{P}_{s}\textrm{P}_{r_{i}}\Gamma_{sr_{i}}\Gamma_{r_{i}d}}{\textrm{P}_{s}\Gamma_{sr_{i}}v_{i}+N_{0}},~~~i=1,2.
\end{equation}
Under the Rayleigh fading channel model, $X_{i} \vert v_{i}$ is exponentially distributed
with parameter $\frac{1}{\mathscr{F}(v_{i})}$ and the probability density function (pdf)
\begin{equation}
f_{X_{i}\vert v_{i}}(x_{i} \vert v_{i})=\dfrac{1}{\mathscr{F}(v_{i})}e^{-\frac{x_{i}}{\mathscr{F}(v_{i})}}U(x_{i}),
\end{equation}
where $U(.)$ is the unit step function. In the high SNR$_{r_{i}d}$ regime (i.e., $\textrm{P}_{r_{i}}\Gamma_{r_{i}d} \gg \textrm{P}_{s}\Gamma_{sr_{i}}$ and $\textrm{P}_{r_{i}}\Gamma_{r_{i}d} \gg N_{0}$, $i=1,2$), (\ref{eqn: 08}) can be simplified to
\begin{eqnarray}
\Omega(\textbf{g}) &\approx& \mathbb{P} \left. \left \lbrace  \dfrac{X_{1} v_{1}+X_{2} v_{2}}{(\dfrac{X_{1}}{\Gamma_{sr_{1}}}+\dfrac{X_{2}}{\Gamma_{sr_{2}}})\dfrac{N_{0}}{\textrm{P}_{s}}} <\gamma_{t} \right \vert \textbf{g} \right \rbrace \\
\label{eqn: 10}&=& \mathbb{P} \left \{ \left(v_{1}-\dfrac{\xi}{\Gamma_{sr_{1}}}\right) X_{1} < \left(\dfrac{\xi}{\Gamma_{sr_{2}}}-v_{2}\right)X_{2}  \Big \vert \textbf{g} \right \},
\end{eqnarray}
where $\xi \triangleq \frac{N_{0}}{\textrm{P}_{s}}\gamma_{t}$.
Depending on the values of $v_{1}$ and $v_{2}$, we have the
following cases:

\underline{\textbf{Case 1:}} $ v_{1} < \frac{\xi}{\Gamma_{sr_{1}}}$
and $v_{2} < \frac{\xi}{\Gamma_{sr_{2}}}$

In this case, (\ref{eqn: 10}) can be written as
\begin{equation}\label{case1}
\Omega(\textbf{g}) = \mathbb{P} \lbrace Z > \phi \vert \textbf{g}  \rbrace \stackrel{(a)}{=} 1,
\end{equation}
where $Z \triangleq \frac{X_{1}}{X_{2}}$ and
\begin{equation}
 \phi \triangleq \dfrac{\frac{\xi}{\Gamma_{sr_{2}}}-v_{2}}{v_{1}-\frac{\xi}{\Gamma_{sr_{1}}}}.
\end{equation}

In the above equations, $(a)$ follows from the fact that for $ v_{1}
< \frac{\xi}{\Gamma_{sr_{1}}}$ and $v_{2} <
\frac{\xi}{\Gamma_{sr_{2}}}$, $\phi$ is negative and this
results in $\mathbb{P} \lbrace Z \leq \phi \vert \textbf{g}
\rbrace=0$.

\underline{\textbf{Case 2:}} $ v_{1} > \frac{\xi}{\Gamma_{sr_{1}}}$
and $v_{2} < \frac{\xi}{\Gamma_{sr_{2}}}$

From (\ref{eqn: 10}), we have
\begin{equation}\label{eqn: 11}
\Omega(\textbf{g}) =  \mathbb{P} \left \lbrace Z < \phi \vert
\textbf{g} \right \rbrace,
\end{equation}
where $\phi$ is positive. It can be shown that the pdf of the random variable
$Z$ conditioned on the vector $\textbf{g}$ is obtained as \cite{JamshidTech2007}
\begin{equation}\label{eqn: 12}
f_{Z\vert \textbf{g}}(z \vert \textbf{g})  = \dfrac{\mathscr{F}(v_{1}) \mathscr{F}(v_{2})}{(\mathscr{F}(v_{2})z+\mathscr{F}(v_{1}))^{2}}U(z).
\end{equation}
Thus, (\ref{eqn: 11}) can be written as
\begin{eqnarray}
\Omega(\textbf{g}) & = & \int_{0}^{\phi} f_{Z\vert \textbf{g}}(z \vert \textbf{g}) dz \\
\label{eqn: 13} & = & \dfrac{\mathscr{F}(v_{2}) \phi}{\mathscr{F}(v_{2})\phi +\mathscr{F}(v_{1})}.
\end{eqnarray}

\underline{\textbf{Case 3:}} $ v_{1} < \frac{\xi}{\Gamma_{sr_{1}}}$ and $v_{2} > \frac{\xi}{\Gamma_{sr_{2}}}$

In this case, (\ref{eqn: 10}) can be written as
\begin{eqnarray}
\Omega(\textbf{g}) & = & \mathbb{P} \lbrace Z > \phi \vert \textbf{g} \rbrace \\
& = & 1- \mathbb{P} \lbrace Z \leq \phi \vert \textbf{g} \rbrace \\
\label{case3}&\stackrel{(a)}{=}&\dfrac{\mathscr{F}(v_{1})}{\mathscr{F}(v_{2})\phi +\mathscr{F}(v_{1})},
\end{eqnarray}
where $\phi$ is positive and $(a)$ comes from (\ref{eqn: 12}) and (\ref{eqn: 13}).

\underline{\textbf{Case 4:}} $ v_{1} > \frac{\xi}{\Gamma_{sr_{1}}}$
and $v_{2} > \frac{\xi}{\Gamma_{sr_{2}}}$

In this case, since $\phi$ is negative, we have
\begin{equation}\label{case4}
\Omega(\textbf{g}) = \mathbb{P} \left \lbrace Z <\phi  \vert \textbf{g} \right \rbrace=0.
\end{equation}

Now, we can use the above results to simplify the outage probability given in (\ref{eqn: 07})
for $\textrm{P}_{r_{i}}\Gamma_{r_{i}d} \gg \textrm{P}_{s}\Gamma_{sr_{i}}$ and
$\textrm{P}_{r_{i}}\Gamma_{r_{i}d} \gg N_{0}$, $i=1,2$. Using (\ref{case1}), (\ref{eqn: 13}),
(\ref{case3}) and (\ref{case4}), we have
\begin{eqnarray}
\textrm{P}_{out} &=&\int_{0}^{\infty} \int_{0}^{\infty} \Omega(\textbf{g}) e^{-v_{1}}e^{-v_{2}}dv_{1}dv_{2}\\
\notag&=& \int_{0}^{\frac{\xi}{\Gamma_{sr_{1}}}}\int_{0}^{\frac{\xi}{\Gamma_{sr_{2}}}}1e^{-v_{1}}e^{-v_{2}}dv_{1}dv_{2}+
\int_{0}^{\frac{\xi}{\Gamma_{sr_{1}}}} \left[\int_{\frac{\xi}{\Gamma_{sr_{2}}}}^{\infty} \dfrac{\mathscr{F}(v_{1})e^{-v_{2}}}{\mathscr{F}(v_{2})\phi +\mathscr{F}(v_{1})}dv_{2}\right]e^{-v_{1}}dv_{1}\\
&+& \int_{\frac{\xi}{\Gamma_{sr_{1}}}}^{\infty} \left[\int_{0}^{\frac{\xi}{\Gamma_{sr_{2}}}} \dfrac{\mathscr{F}(v_{2})\phi e^{-v_{2}}}{\mathscr{F}(v_{2})\phi +\mathscr{F}(v_{1})}dv_{2}\right]e^{-v_{1}}dv_{1}+\int_{\frac{\xi}{\Gamma_{sr_{1}}}}^{\infty}\int_{\frac{\xi}{\Gamma_{sr_{2}}}}^{\infty}0e^{-v_{1}}e^{-v_{2}}dv_{1}dv_{2}\\
\notag & = & \left(1-e^{-\frac{\xi}{\Gamma_{sr_{1}}}}\right)\left(1-e^{-\frac{\xi}{\Gamma_{sr_{2}}}}\right)+
\int_{0}^{\frac{\xi}{\Gamma_{sr_{1}}}} \left[\int_{\frac{\xi}{\Gamma_{sr_{2}}}}^{\infty} \dfrac{e^{-v_{2}}}{\mathscr{F}(v_{2})\phi +\mathscr{F}(v_{1})}dv_{2}\right]\mathscr{F}(v_{1})e^{-v_{1}}dv_{1}\\
\label{eqn: 15}&+&  \int_{0}^{\frac{\xi}{\Gamma_{sr_{2}}}}\left[\int_{\frac{\xi}{\Gamma_{sr_{1}}}}^{\infty} \dfrac{\phi e^{-v_{1}}}{\mathscr{F}(v_{2})\phi +\mathscr{F}(v_{1})}dv_{1}\right]\mathscr{F}(v_{2})e^{-v_{2}}dv_{2}.
\end{eqnarray}
Now, the main objective is to find the optimum scaling factor $\alpha_{i}$ (or equivalently
$\mathscr{F}(v_{i})$ defined in (\ref{eq: 09})) that minimizes the outage probability obtained
in (\ref{eqn: 15}). Since for $ v_{1} < \frac{\xi}{\Gamma_{sr_{1}}}$ and $v_{2} > \frac{\xi}{\Gamma_{sr_{2}}}$,
$\phi \triangleq \frac{\frac{\xi}{\Gamma_{sr_{2}}}-v_{2}}{v_{1}-\frac{\xi}{\Gamma_{sr_{1}}}} >0$, it
is concluded that the second term in (\ref{eqn: 15}) is a nonnegative value. With a similar argument,
the third term in (\ref{eqn: 15}) is nonnegative as well. Also, we use the fact that
$\xi \triangleq \frac{N_{0}}{\textrm{P}_{s}}\gamma_{t} >0$. Hence, to minimize (\ref{eqn: 15}),
it is sufficient to have
\begin{eqnarray}
\label{setzero}\int_{0}^{\frac{\xi}{\Gamma_{sr_{1}}}} \left[\int_{\frac{\xi}{\Gamma_{sr_{2}}}}^{\infty} \dfrac{e^{-v_{2}}}{\mathscr{F}(v_{2})\phi+\mathscr{F}(v_{1})}dv_{2}\right]
\mathscr{F}(v_{1})e^{-v_{1}}dv_{1}&=&0,\\
\label{setzero1}\int_{0}^{\frac{\xi}{\Gamma_{sr_{2}}}}\left[\int_{\frac{\xi}{\Gamma_{sr_{1}}}}^{\infty} \dfrac{\phi e^{-v_{1}}}{\mathscr{F}(v_{2})\phi +\mathscr{F}(v_{1})}dv_{1}\right]\mathscr{F}(v_{2})e^{-v_{2}}dv_{2}&=&0.
\end{eqnarray}
Noting that $\int_{\frac{\xi}{\Gamma_{sr_{2}}}}^{\infty} \frac{e^{-v_{2}}}{\mathscr{F}(v_{2})\phi+\mathscr{F}(v_{1})}dv_{2}$
and $\int_{\frac{\xi}{\Gamma_{sr_{1}}}}^{\infty} \dfrac{\phi e^{-v_{1}}}{\mathscr{F}(v_{2})\phi +\mathscr{F}(v_{1})}dv_{1}$
are positive, the optimum function $\mathscr{F}(v_{i})$ that satisfies (\ref{setzero}) and (\ref{setzero1})
(or equivalently minimizes (\ref{eqn: 15})) is
\begin{equation}\label{result01}
\mathscr{F}(v_{i})=0,~~ \textrm{for}~ 0 < v_{i} < \frac{\xi}{\Gamma_{sr_{i}}},~~~i=1,2
\end{equation}
In this case,
\begin{eqnarray}
\textrm{min}~ \textrm{P}_{out}&=&\left(1-e^{-\frac{\xi}{\Gamma_{sr_{1}}}}\right)\left(1-e^{-\frac{\xi}{\Gamma_{sr_{2}}}}\right) \\
&\stackrel{(a)}{=}& \left(1-e^{-\frac{\gamma_{t}}{\textrm{SNR}_{sr_{1}}}}\right)\left(1-e^{-\frac{\gamma_{t}}{\textrm{SNR}_{sr_{2}}}}\right),
\end{eqnarray}
where $(a)$ comes from $\xi \triangleq \frac{N_{0}}{\textrm{P}_{s}}\gamma_{t}$ and SNR$_{sr_{i}} \triangleq \frac{\textrm{P}_{s}}{N_{0}}\Gamma_{sr_{i}}$.
Using (\ref{eq: 09}) and (\ref{result01}), we come up with
the following result:
\begin{equation}
\hat{\alpha}_{i}=0,~~~~0 < g_{sr_{i}} <  \xi, ~~i=1,2,
\end{equation}
where $\hat{\alpha}_{i}$ is the optimum scaling factor that minimizes the outage probability.
In other words, the relay remains silent if its channel gain with the source
is less than the threshold level $\xi$.
Note that we are not concerned with the computation of the exact values of the scaling factors
at the relay nodes, i.e., $\hat{\alpha}_{i}$ for $v_{i} \geq \frac{\xi}{\Gamma_{sr_{i}}}$, as long as $\textrm{P}_{r_{i}}\Gamma_{r_{i}d} \gg \textrm{P}_{s}\Gamma_{sr_{i}}$ and $\textrm{P}_{r_{i}}\Gamma_{r_{i}d} \gg N_{0}$, $i=1,2$.

\textbf{Remark 1-} The case of high SNR$_{sr_{i}}$ is similar to the traditional STC problem. Therefore, the optimum power allocation policy for
relay stations is the full power transmission.

\section{Numerical Results}\label{simulation}
In this section, we present some Monte-Carlo simulation results to evaluate the impact
of $\alpha_{i}$ on the system performance.
We use the propagation model with \textit{intermediate path-loss condition} (i.e., terrain
type B for suburban, above roof top (ART) to below roof top (BRT)) provided by
WiMAX Forum and IEEE 802.16j relay Working Group (WG) \cite{IEEE802.16j}, i.e.,
\begin{eqnarray}\nonumber
    \Upsilon_{ij} \triangleq
\left\{\begin{array}{ll}
20 \log_{10} \left(\dfrac{4\pi d_{ij}}{\lambda} \right)   ,    & d_{ij} \leq d^{'}_{0} \\
K+10\beta \log_{10} \left( \dfrac{d_{ij}}{d_{0}} \right) +\Delta \Upsilon_{f}+\Delta \Upsilon_{h_{t}} ,  &  d_{ij} > d^{'}_{0},
\end{array} \right.
\end{eqnarray}
where $\Upsilon_{ij} \triangleq -10 \log_{10} \Gamma_{ij}$ is the path loss between nodes $i$ and $j$ expressed in dB, and
\begin{itemize}
\item $d_{ij}$ is the distance between nodes $i$ and $j$,
\item $\lambda$ is the wavelength,
\item $\beta$ is the path loss exponent,
\item $d_{0}$ is the reference distance,
\item $d_{0}^{'} \triangleq d_{0}10^{-\frac{\Delta \Upsilon_{f}+\Delta \Upsilon_{h_{t}}}{10\beta}}$
\item $\Delta \Upsilon_{f}$ is the correction factor for the carrier frequency $f$,
\item $\Delta \Upsilon_{h_{t}}$ is the correction factor for the RS/MS BRT antenna height $h_{t}$,
\item $K \triangleq 20 \log_{10} \left(\dfrac{4\pi d^{'}_{0}}{\lambda}
\right)$.
\end{itemize}

For this model, $\beta$, $\Delta \Upsilon_{f}$ and $\Delta \Upsilon_{h_{t}}$ are given as \cite{IEEE802.16j}
\begin{equation}
 \beta \triangleq a-bh_{b}+\dfrac{c}{h_{b}},  \notag
\end{equation}
\begin{equation}
 \Delta \Upsilon_{f} \triangleq 6\log_{10} \left( \dfrac{f(\textrm{MHz})}{2000} \right),    \notag
\end{equation}
\begin{eqnarray}\nonumber
    \Delta \Upsilon_{h_{t}} \triangleq
\left\{\begin{array}{ll}
-10 \log_{10} \left(\dfrac{h_{t}}{3} \right)   ,    & h_{t} \leq 3~\textrm{m} \\
-20 \log_{10} \left(\dfrac{h_{t}}{3} \right)   ,    & h_{t} > 3~\textrm{m},
\end{array} \right.
\end{eqnarray}
respectively, where $h_{b}$ represents the height of BS/RS ART antenna.
Using the system parameters listed in Table II, the desired SNR$_{sr_{i}}$ and SNR$_{r_{i}d}$ are obtained
through adjusting $\frac{\textrm{P}_{s}}{N_{0}}$ and $\frac{\textrm{P}_{r_{i}}}{N_{0}}$, respectively.
\begin{table}
   \label{table2}
\caption{System Evaluation Parameters \cite{IEEE802.16j}}
\centering
  \begin{tabular}{|c|c|}
  \hline
   \textbf{Parameters} & \textbf{Values} \\
  \hline
   Carrier frequency &  2.4 GHz \\
  \hline
   SNR$_{r_{i}d}$, $i=1,2$ &  $0-40$ dB \\
  \hline
  $d_{0}$   & 100 m\\
  \hline
  Antenna height   & BS=32 m, RS=15 m, MS=1.5 m\\
  \hline
  Distance between S and RS   &  $d_{sr_{1}}=5$ km, $d_{sr_{2}}=8$ km\\
  \hline
  Distance between RS and D & $d_{r_{1}d}=2$ km, $d_{r_{2}d}=1$ km\\
   \hline
  Path loss exponent parameters & $a=4$, $b=0.0065$, $c=17.1$\\
   \hline
  \end{tabular}
\end{table}

We consider some forms of the scaling factors $\alpha_{i}$ in terms of $g_{sr_{i}}$.
In the \textit{full power} scheme, each relay retransmits its received signal with
full power (i.e., $\alpha_{i}=1$), regardless of the value of $g_{sr_{i}}$. For the
\textit{on-off power} scheme, RS$_{i}$ with $g_{sr_{i}}$ above certain threshold
transmits at full power, otherwise it remains silent, i.e., $\alpha_{i}=0$. We also
consider the \textit{piecewise linear function} scheme, in which the scaling
factor $\alpha_{i}$ is defined as
\begin{eqnarray}
    \alpha_{i} \triangleq
\left\{\begin{array}{ll}
0,    &  g_{sr_{i}} < \tau_{1} \\
\dfrac{1}{\tau_{2}-\tau_{1}}g_{sr_{i}}-\dfrac{\tau_{1}}{\tau_{2}-\tau_{1}}  ,  &  \tau_{1} \leq g_{sr_{i}} < \tau_{2} \\
1,  &  g_{sr_{i}} \geq \tau_{2},
\end{array} \right.
\end{eqnarray}
where the optimum values of $\tau_{1}$ and $\tau_{2}$ are obtained using an exhaustive search
to minimize the outage probability.
This function can approximate a large class of appropriate
functions\footnote{The optimum function should have small values when the channel is weak and
should increase up to 1 as the channel gain increases.} by adjusting $\tau_{1}$ and $\tau_{2}$.
In addition, we use the ML decoder at D to fully exploit the diversity advantage of the proposed scheme.

\begin{figure}[t]
\centerline{\psfig{figure=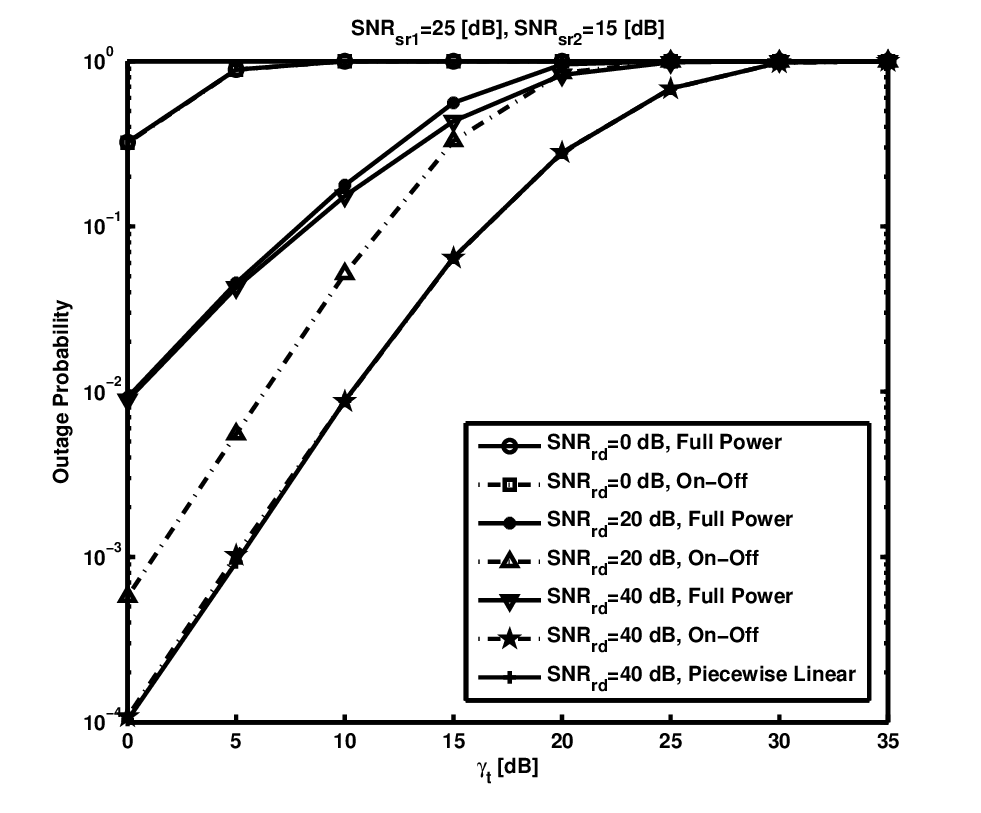,width=4.05in}}
\caption{Outage probability versus $\gamma_{t}$ for different strategies.}
\label{fig: total1}
\end{figure}

Fig. \ref{fig: total1} compares the outage probability of the aforementioned
power schemes versus $\gamma_{t}$ for
different values of SNR$_{r_{1}d}=$SNR$_{r_{2}d}\triangleq$SNR$_{rd}$.
The results are obtained by averaging over $10^{6}$ channel realizations.
We can observe a significant performance improvement for the on-off power
scheme in comparison with the full power scenario, as SNR$_{rd}$ increases.
For instance, $10$ dB gain is obtained using the on-off power strategy for SNR$_{rd}=40$ dB and
$\textrm{P}_{out}=10^{-2}$ with respect to the full power scheme. Also,
it can be seen that the optimized piecewise linear function performs close
to the on-off scheme. Table III shows the
optimum threshold level of the on-off power scheme expressed in dB for different values of
$\gamma_{t}$ and SNR$_{rd}$, and SNR$_{sr_{1}}=$SNR$_{sr_{2}}=25$ dB. It is
interesting to note that the optimum threshold values obtained by exhaustive
search are close to $\xi$ (dB) $=\gamma_{t}$ (dB) $-$ SNR$_{sr_{i}}$ (dB)
obtained in Section \ref{analysis}.

\begin{table*}
   \label{table3}
\caption{Optimum Threshold Level for the On-Off Power Scheme}
\centering
  \begin{tabular}{|c|c|c|c|c|c|}
  \hline
   & SNR$_{rd}=0$  dB & SNR$_{rd}=10$ dB & SNR$_{rd}=20$ dB & SNR$_{rd}=30$ dB & SNR$_{rd}=40$ dB \\
  \hline
   $\gamma_{t}=0$ dB &  -25  & -25  & -25 & -25  & -25  \\
  \hline
  $\gamma_{t}=10$ dB &  -13  & -15  & -15 & -15 & -15 \\
  \hline
  $\gamma_{t}=20$ dB &  0   & -1.5 & -5  & -5 & -5   \\
  \hline
  $\gamma_{t}=30$ dB &  0  & 0  & 0  & 0 & 0\\
  \hline
  \end{tabular}
\end{table*}

\begin{figure}[t]
\centerline{\psfig{figure=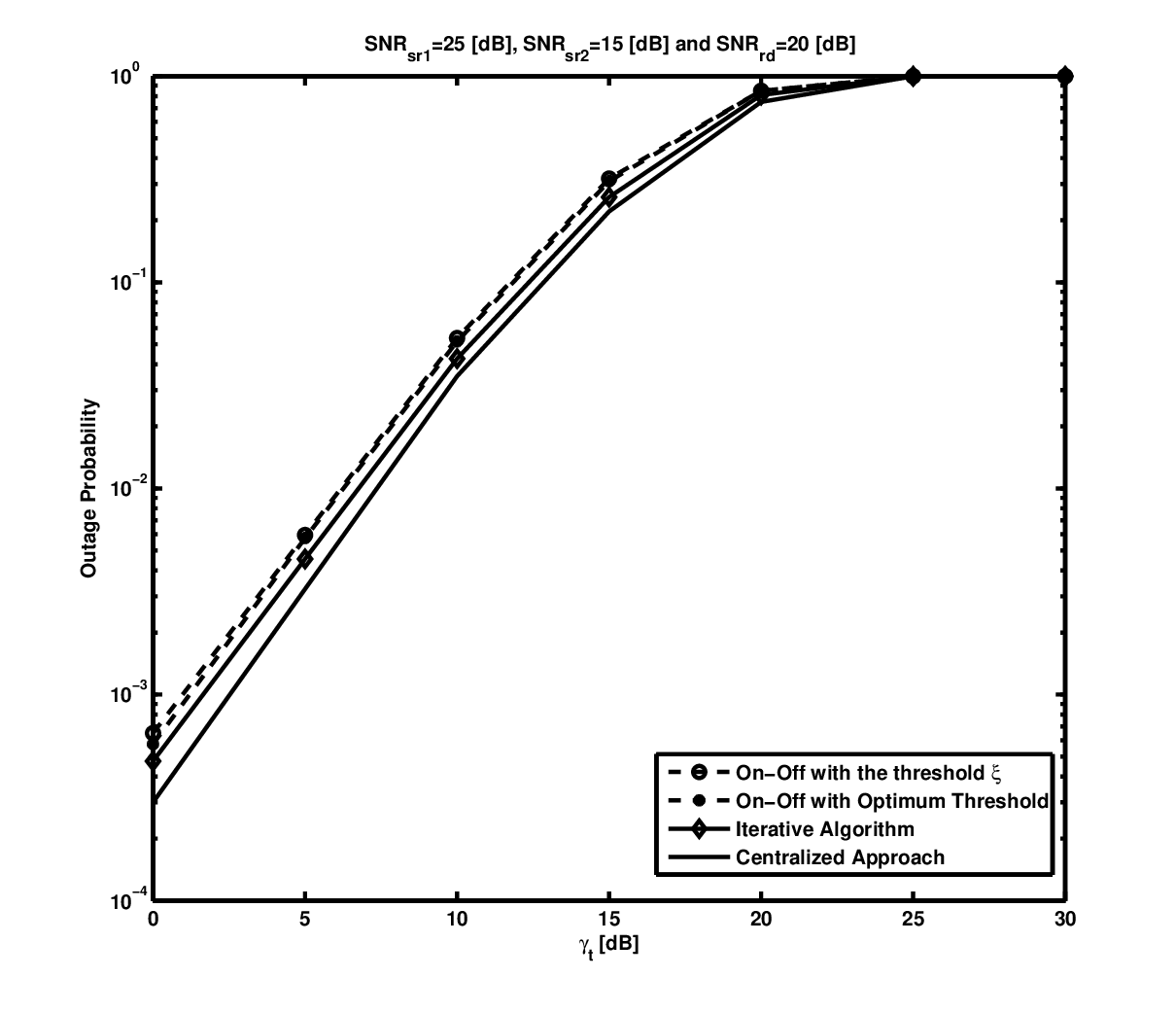,width=4.54in}}
\caption{Outage probability versus $\gamma_{t}$ for different power schemes.}
\label{fig: iterative}
\end{figure}

In Fig. \ref{fig: iterative}, we plot the outage probabilities of the centralized scheme, iterative algorithm and the on-off power allocation
strategy with the threshold $\xi$ obtained in Section \ref{analysis}. In the centralized approach, the optimum $\alpha_{1}$ and $\alpha_{2}$
are obtained through an exhaustive search to minimize the outage probability. In the iterative scheme, the algorithm starts from an initial value of $\alpha_{1}$ for every $g_{sr_{i}}$. Then, for each value of $g_{sr{2}}$, $\alpha_{2}$ is obtained such that the outage probability is
minimized. This process continues until the algorithm converges. The results in Fig. \ref{fig: iterative} show that the difference
in performance between the proposed on-off power scheme and the centralized approach is small. Therefore, the on-off scheme obtained by the asymptotic
analysis in Section \ref{analysis} is almost optimal.

\begin{figure}[t]
\centerline{\psfig{figure=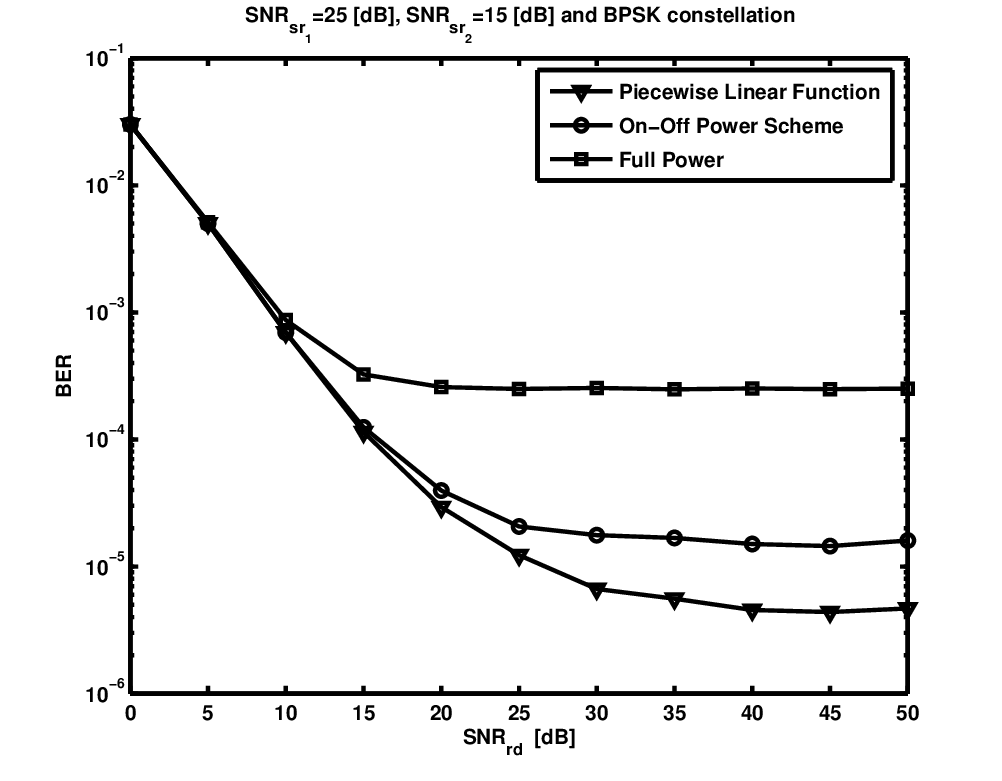,width=4.04in}}
\caption{Uncoded BER versus SNR$_{rd}$ for different power schemes.}
\label{fig: BERnew}
\end{figure}

Fig. \ref{fig: BERnew} plots the uncoded BER
versus SNR$_{rd}$ for different power schemes and BPSK
constellation. We observe an error floor in the figure
for the high SNR$_{rd}$. This can be attributed to the amplified
noise received at the destination through the R $\rightarrow$ D link. In
this case, the performance is governed by SNR$_{sr_{i}}$, $i=1,2$.

\begin{figure}[t]
\centerline{\psfig{figure=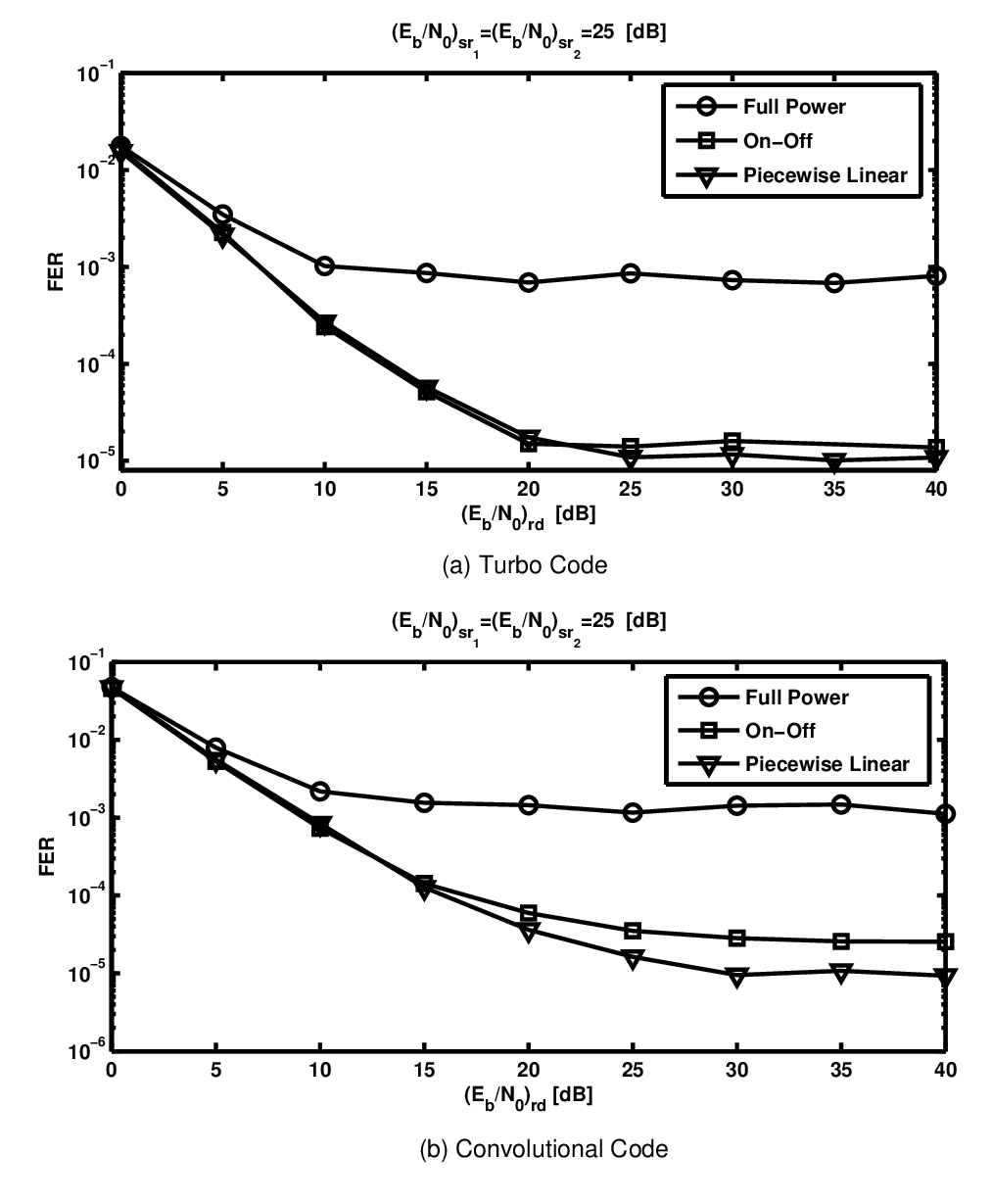,width=3.5in}}
\caption{FER versus $\left(E_{b}/N_{0}\right)_{rd}$ for different power schemes and for a) turbo code with the rate $1/3$  b) convolutional code with the rate $1/2$.}
\label{fig: FER}
\end{figure}

It should be noted that by using the on-off power allocation strategy at RSs,
we can save the energy in the silent mode. However, simulation results implies that the
power saving gain in the on-off power scheme is negligible. This is due to the
fact that for mid-SNR$_{sr_{i}}$ region, $\textrm{P}_{_{ON}} \triangleq \textrm{Pr}\{g_{sr_{i}} >g_{Th}\} \geq 0.95$,
where $g_{Th}$ is the optimum threshold level that minimizes the BER.

Fig. \ref{fig: FER}-a illustrates the frame-error rate (FER)
versus $\left(E_{b}/N_{0} \right)_{rd}$, energy per bit to noise density ratio, using the standard turbo code
\cite{BerrouICC93} with the rate of $1/3$ and for different
functions of $\alpha_{i}$, BPSK constellation and the frame length
of $957$ bits. Also, Fig. \ref{fig: FER}-b illustrates the FER
versus $\left(E_{b}/N_{0} \right)_{rd}$ using the convolutional code (introduced in
section $8.4.9.2.1$ of the IEEE 802.16e standard \cite{IEEE802.16e})
with the rate of $0.5$ and the frame length of $576$ bits. Compared
to the full power scheme, the simulation results show a significant
improvement in the system performance with the on-off power scheme.
Also, it is observed from the simulation that the trend of the FER
for higher order constellations is similar to the BPSK case
\cite{JamshidTech2007}.

Up to now, we have shown that the near-optimal power allocation in
each relay in the AF mode is the threshold-based on-off power
scheme. To further improve the system performance when the quality of S$\rightarrow$RS$_{i}$ is good, the
detect-and-forward (DF) scheme is suggested to eliminate the
amplified noise at the RS. A \textit{hybrid threshold-based AF/DF} scheme characterized by two
threshold levels $T_{1}$ and $T_{2}$ is described as follows:

\begin{itemize}
\item [1-] For $g_{sr_{i}} \leq T_{2}$, RS$_{i}$ performs the on-off power scheme developed in Section \ref{analysis}.
\item [2-] For $g_{sr_{i}} > T_{2}$, RS$_{i}$ detects $x[k]$, $k \in \{1,2 \}$, from the received signal and forwards the detected
symbol $\hat{x}[k]$ to the destination.
\end{itemize}

\begin{figure}[t]
\centerline{\psfig{figure=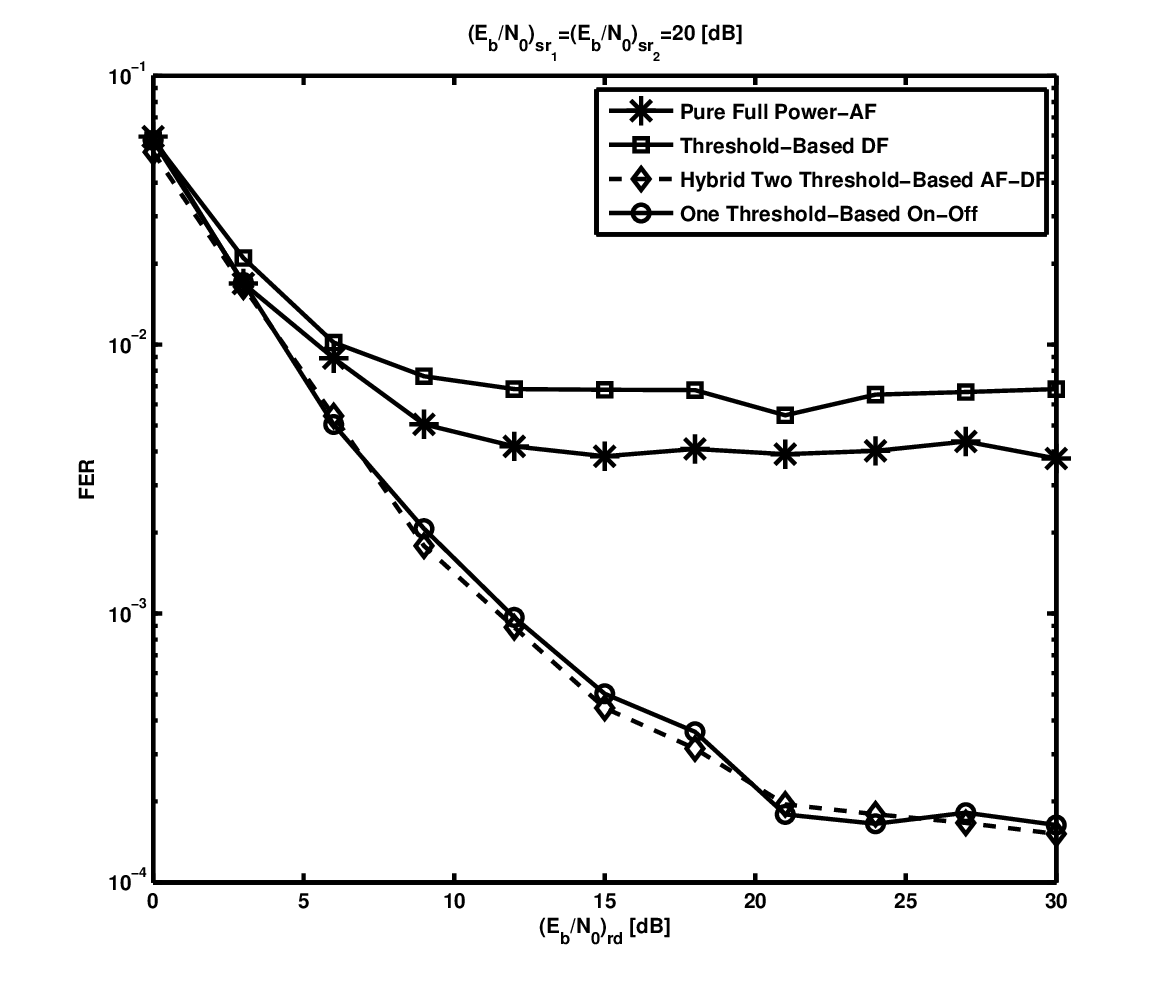,width=4.04in}}
\caption{FER versus $\left(E_{b}/N_{0} \right)_{rd}$ for different transmission strategies and for convolutional code with the rate $1/2$, $\left(E_{b}/N_{0} \right)_{sr}=20~$dB and for QPSK.}
\label{fig: DF1}
\end{figure}

Fig. \ref{fig: DF1} compares the FER of different transmission
strategies in the RSs for the convolutional code with QPSK
modulation and the frame length of $576$ bits. In the pure full
power scheme, independent of the channel gain $g_{sr_{i}}$, each
RS$_{i}$ transmits with full power all the time. For completeness,
we also consider the threshold-based DF scheme, in which for
$g_{sr_{i}}$ less than a prescribed threshold, RS$_{i}$
remains silent, otherwise it switches to the DF mode. It is observed
that the performance of the hybrid threshold-based AF/DF scheme is
the same as the threshold-based AF on-off power scheme and both of them
are better than the pure full power and the threshold-based DF schemes.

\section{Impact of Channel Estimation Error}\label{robust}
Throughout this paper, it is assumed that perfect channel knowledge
is available at the receiver. In practice, however, the channel
estimation at the receiver is often imperfect. Thus, the performance
of the system is degraded due to the channel estimation
error\footnote{Such a problem has been first studied by Alamouti \cite{AlamoutiJSAC0898} and Tarokh
\cite{TarokhITC0299} in STC-based systems. This line of work is
further extended for different structures of the receiver (See
\cite{HoteitIEEProc2005} and its references).}. In this section, we
evaluate the FER degradation of the proposed scheme due to the
imperfect channel estimation.
To handle that, a pilot-based channel estimation technique is
assumed, where pilot sequences are inserted at the beginning of each
block. It is assumed that the pilot sequences have the same energy
as data symbols. In this case, the channel estimation process is
performed at the beginning of each block and the estimated value
remains constant over the block. This assumption is valid under the
frequency-flat Rayleigh fading channel model, where the multipath
channel coefficients change slowly and are considered constant. Due
to the imperfect channel estimation, the channel fading coefficients
can be expressed as
\begin{eqnarray}
\hat{h}_{sr_{i}}&=&h_{sr_{i}}+e_{sr_{i}},\\
\hat{h}_{r_{i}d}&=&h_{r_{i}d}+e_{r_{i}d},
\end{eqnarray}
for $i=1,2$, where $e_{sr_{i}}$ and $e_{r_{i}d}$ represent the
channel estimation errors which are modeled as independent complex
Gaussian random variables with zero means and variances
$\vartheta_{sr_{i}}$ and $\vartheta_{r_{i}d}$, respectively. We assume that the
channel estimation error has the variance equal to the
inverse of the signal-to-noise ratio of the corresponding link
\cite{TarokhITC0299, BuehrerVTC2002, MavaresVTC2003}, i.e., $\vartheta_{sr_{i}}=1/\textrm{SNR}_{sr_{i}}$
and $\vartheta_{r_{i}d}=1/\textrm{SNR}_{r_{i}d}$.

\begin{figure}[t]
\centerline{\psfig{figure=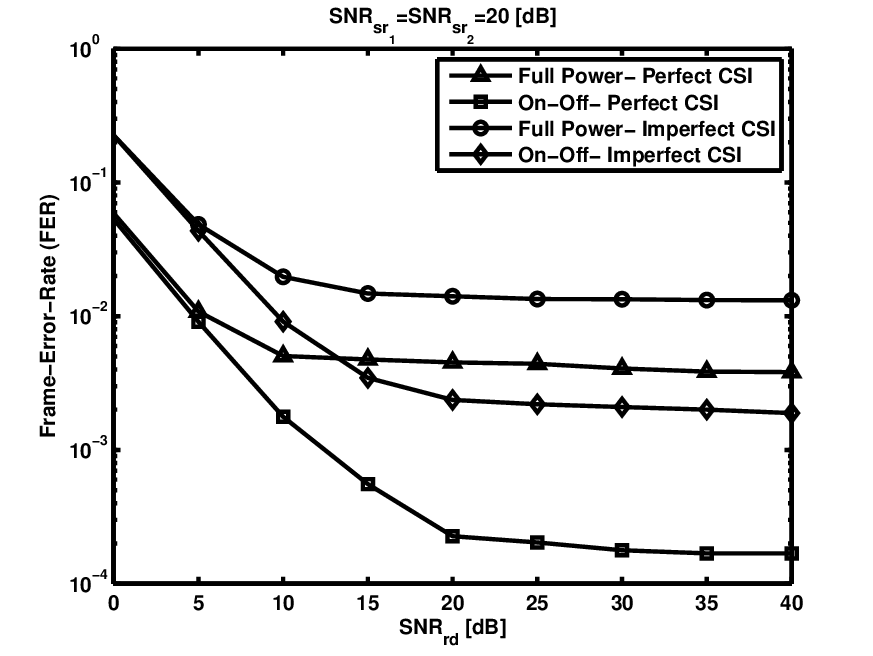,width=4.04in}}
\caption{FER versus SNR$_{rd}$ for the full power and the on-off power allocation schemes for perfect and imperfect CSI.}
\label{fig: robustness}
\end{figure}

Fig. \ref{fig: robustness} shows the FER performance of the full
power and the on-off power schemes versus SNR$_{rd}$ for perfect and
imperfect CSI situations. We observe that for the low SNR$_{rd}$, the
robustness of the on-off power scheme is close to the
full power case. However, for high SNR$_{rd}$ regime, the full
power scheme is more robust to channel estimation errors than the
on-off power scheme. This is due to the fact that in the on-off
scheme, each relay $i$ requires CSI of the link S $\rightarrow$
RS$_{i}$ for scaling factor computation. It is obvious that in this
case, imperfect CSI further deteriorates the performance. Note that
at low SNR$_{rd}$, the performance is governed by the second link.
In this case, CSI of the link RS$_{i}$ $\rightarrow$ D becomes
equally important in both schemes.

\section{Conclusion}\label{conclusion1}
In this paper, a non-regenerative dual-hop wireless system based on
a distributed space-time coding strategy is considered. It is
assumed that each relay forwards an appropriately scaled space-time
coded version of its received signal. In the high SNR$_{r_{i}d}$, it
has been shown that the optimum power allocation strategy in each
RS$_{i}$ which minimizes the outage probability is to remain
silent, if $g_{sr_{i}}$ is less than a prespecified threshold level.
Simulation results show that the threshold-based on-off power scheme
performs close to optimum for finite SNR values. Numerical results
demonstrate a dramatic improvement in system performance as compared
to the case that the relay stations
forward their received signals with full power. The practical
advantage of using this protocol is that it does not alter the
Alamouti decoder at the destination, while it slightly modifies the
relay operation.

\section*{Acknowledgment}
The authors would like to thank S. A. Motahari, V. Pourahmadi and S. A. Ahmadzadeh of CST Lab. for the helpful discussions.


\end{document}